%% file: main.tex
\documentclass[aps,twocolumn, 
onecolumnshowpacs,superscriptaddress,tightenlines,10pt,longbibliography]{revtex4-2} 
\usepackage{graphicx}
\usepackage{color,soul}
\usepackage{xcolor}
\usepackage{braket}
\usepackage{hyperref}
\usepackage{comment}
\usepackage{amsmath}
\usepackage{appendix}
\usepackage{physics}
\usepackage[separate-uncertainty=true]{siunitx}
\usepackage{microtype}
\usepackage{lmodern}
\usepackage{import}
\usepackage{xifthen}
\usepackage{pdfpages}
\usepackage{transparent}
\graphicspath{{figures/}}
\usepackage{mwe}

\def\g2{g^{(2)}(0)}

\makeatletter
\newcommand*{\defeq}{\mathrel{\rlap{%
                     \raisebox{0.3ex}{$\m@th\cdot$}}%
                     \raisebox{-0.3ex}{$\m@th\cdot$}}%
                     =}
\makeatother

\makeatletter
    \renewcommand\@make@capt@title[2]{%
     \@ifx@empty\float@link{\@firstofone}{\expandafter\href\expandafter{\float@link}}%
      {\textbf{#1}}\@caption@fignum@sep#2\quad}%
    \makeatother
   
\UseRawInputEncoding
\usepackage{lipsum}
\makeatletter
\AtBeginDocument{\let\LS@rot\@undefined}
\makeatother

\begin{document}
\title{Near-unity indistinguishability of single photons emitted from \\ dissimilar and independent atomic quantum nodes}

\author{F\'elix Hoffet}
\email{felix.hoffet@icfo.eu}
\affiliation{ICFO -- Institut de Ciencies Fotoniques, The Barcelona Institute of Science and Technology, Spain}

\author{Jan Lowinski}
\affiliation{ICFO -- Institut de Ciencies Fotoniques, The Barcelona Institute of Science and Technology, Spain}

\author{Lukas Heller}
\affiliation{ICFO -- Institut de Ciencies Fotoniques, The Barcelona Institute of Science and Technology, Spain}

\author{Auxiliadora Padr\'on-Brito}
\affiliation{ICFO -- Institut de Ciencies Fotoniques, The Barcelona Institute of Science and Technology, Spain}

\author{Hugues de Riedmatten}
\email{hugues.deriedmatten@icfo.eu}
\affiliation{ICFO -- Institut de Ciencies Fotoniques, The Barcelona Institute of Science and Technology, Spain}
\affiliation{ICREA -- Instituci\'o Catalana de Recerca i Estudis Avan\c cats, 08015 Barcelona, Spain}

\begin{abstract}
Generating indistinguishable photons from independent nodes is an important challenge for the development of quantum networks. In this work, we demonstrate the generation of highly indistinguishable single photons from two dissimilar atomic quantum nodes. One node is based on a fully blockaded cold Rydberg ensemble and generates on-demand single photons. The other node is a quantum repeater node based on a DLCZ quantum memory and emits heralded single photons after a controllable memory time that is used to synchronize the two sources. We demonstrate an indistinguishability of $\SI{94.6 \pm 5.2}{\percent}$ for a temporal window including $\SI{90}{\percent}$ of the photons.
This advancement opens new possibilities for interconnecting quantum repeater and processing nodes with high fidelity Bell-state measurement without sacrificing its efficiency.
\end{abstract}

\maketitle

\section{Introduction}
\label{sec:intro}

Single photons play a crucial role in quantum technologies. 
Since their quantum state is little affected by the transmission, they are well-suited as information carriers to connect and entangle distant quantum nodes \cite{Kimble2008a}.
In a quantum network, these quantum nodes serve specific roles, dedicated either to information processing, communication, or sensing \cite{Wehner2018}. 
Consequently, they utilize various physical systems, often employing photons with differing frequencies, bandwidths, and waveforms.
A major challenge is to interface these different nodes and distribute entanglement between them, despite their inherent differences. 
This can be achieved by interfering single photons emitted from these nodes, but a high degree of indistinguishability between the photons is required \cite{Sangouard2011}.

The indistinguishability $\eta$ between two photons can be inferred by measuring the visibility $V$ of a Hong-Ou-Mandel interference \cite{Hong1987}.
When two photons impinge on a beamsplitter, the Hong-Ou-Mandel effect states that if they are indistinguishible, they will bunch at the output, leading to a suppression of coincidence events relative to the distinguishable scenario.
This two-photon interference is also at the heart of photonic Bell state measurements, which enable quantum teleportation and the generation of entanglement between remote systems \cite{Pirandola2015}.

High indistinguishabilities have been demonstrated for consecutive photons emitted from a single source, including quantum dots \cite{Ding2016, Cogan2023, Huber2017, Coste2022}, single rare-earth ions \cite{Ourari2023}, Rydberg ensembles \cite{Ornelas-Huerta2020,Shi2022}, spontaneous parametric down conversion \cite{Bruno2014, Xiong2016} and single atoms in cavities \cite{Nisbet-Jones2011}. 
However, it is important to note that these results do not guarantee that independent photons from two of these sources would be indistinguishable.
Ensuring the short-term stability of photon emission is different from achieving mode-matching between independent sources over long periods, the latter being a more challenging task.
Recently, a handful of experiments have demonstrated two-photon interference from independent sources \cite{Felinto2006, Sipahigil2014, Samara2021, Koong2022, You2022, Waltrich2023, Stolk2022a, Duquennoy2022, Zhai2022, Zhang2022b}, sometimes allowing the generation of remote entanglement between nodes based on the same platform \cite{Chou2005, Krutyanskiy2023, vanLeent2022, Zhang2022, Yu2020a}. 

\begin{figure*}[t]
    \begin{minipage}[c]{0.9\linewidth}
   \def\svgwidth{\columnwidth}
   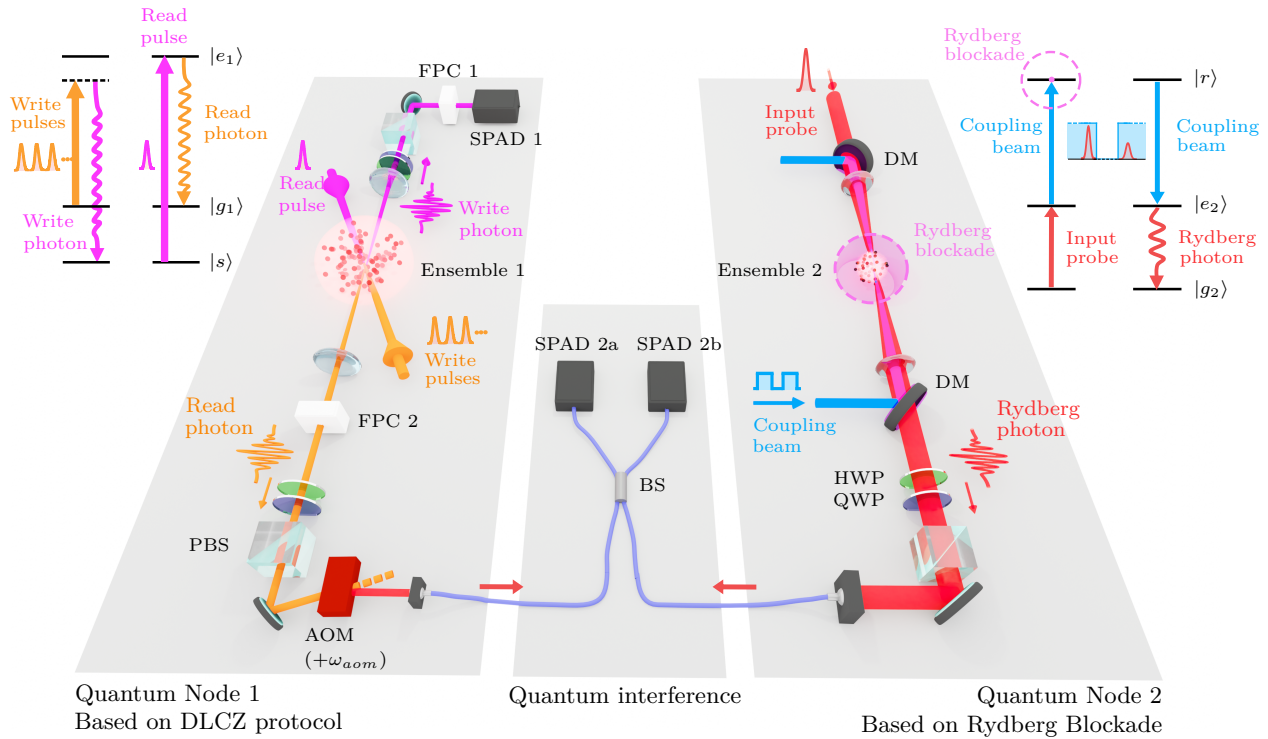
    \end{minipage}
    \caption{Schematic view of the experiment.
    Two quantum nodes generate independent single photons that are sent to a measurement setup, where they interfere.
    Quantum node 1 is based on the DLCZ protocol, where the detection of a write photon in a detector (SPAD1) heralds a collective excitation.
    This excitation is later mapped to a waveshape-tunable single photon by applying a read pulse.
    At the same time, Node 2 generates a single photon by means of dynamical Rydberg-EIT in a Rydberg-blockaded ensemble.
    The photon generated in Node 1 is frequency shifted by a single-pass acousto-optic modulator (AOM) to match the one generated in Node 2.
    Both photons are then sent to a Hong-Ou-Mandel setup consisting of two detectors (SPAD2a(b)) and a beam-splitter (BS).
    Coincidences are recorded in the case when both photons are set to be overlapping (indistinguishable case), and temporally separated (distinguishable case).
    DM:~Dichroic mirror; PBS:~Polarizing beam-splitter; FPC:~Fabry-Perot cavity; HWP(QWP): Half-(quarter-)wave plate}
    \label{fig:setup}
\end{figure*}

However, most of these realizations had to discard a large amount of their photon detections to keep the two-photon indistinguishability sufficiently high.
This is done by selecting a temporal detection window much shorter than the photon duration.
In quantum communication, this selection is detrimental as it reduces the entanglement distribution rate between the two nodes.
Up to now, it is still a challenge to interface independant nodes with high indistinguishability, without discarding any photon detections.

Additionally, interfacing hybrid systems adds another layer of complexity since their generated photons usually have different properties (e.g. frequencies, bandwidths, temporal modes), making them distinguishable. 
Hong-Ou-Mandel interferences between a single-photon source and a classical source such as a laser \cite{Bennett2009}, or the Sun \cite{Deng2019}, have been demonstrated before, albeit with an overall visibility below the classical limit of 50\%.
To our knowledge, only one experiment measured the indistinguishability of photons emitted from dissimilar and independent matter quantum nodes \cite{Craddock2019}, but again the overall interference visibility couldn't surpass the classical limit due to temporal-mode mismatch.

Platforms currently investigated as quantum network nodes include laser cooled atoms and ions, as well as solid-state systems. 
Cold atomic-ensembles provide a versatile platform that enables several functionalities. 
Ground state collective excitations (spin waves) allow for the realization of long-lived and efficient quantum memories that can be used as quantum repeater nodes, using the Duan-Lukin-Cirac-Zoller (DLCZ) scheme \cite{Duan2001,Sangouard2011, Yu2020a, Felinto2006}. 
Additionally, atomic ensembles enable quantum processing nodes  \cite{Firstenberg2016,Paredes-Barato2014, Adams2019, Busche2017, Tiarks2019, Stolz2022} and deterministic quantum light sources \cite{Ornelas-Huerta2020,Shi2022,Lowinski2024} using the Rydberg blockade effect. 

In this work, we demonstrate the generation of highly indistinguishable photons from two dissimilar cold atomic quantum nodes. 
Node 1 relies on the probabilistic and heralded generation of spin waves following the DLCZ protocol, later converted into a single photon after a controllable storage time used to synchronize the nodes. 
Triggered by Node 1, Node 2 then generates single photons in a deterministic fashion using Rydberg electromagnetically-induced transparency \cite{Pritchard2010,Petrosyan2011}, and the two photons are combined at a beam splitter.  
We achieve a Hong-Ou-Mandel visibility of $V$=\SI{80.2 \pm 3.6}{\percent} leading to an indistinguishability of  $\eta$=\SI{89 \pm 4.9}{\percent} between the independent photons, without discarding any detection. 
By considering a temporal window containing \SI{75}{\percent} of the photon counts, the visibility increases to ${V= \SI{93.2 \pm 2.3}{\percent}}$. 
This experiment opens up new possibilities for connecting remote dissimilar quantum nodes.

\section{Experimental details}
\label{sec:setup}

Our two quantum nodes are based on ensembles of cold $^{87}$Rb atoms.
They are located in the same laboratory on two optical tables separated by \SI{3}{\meter} (see \autoref{fig:setup}).
We first start with the individual description of each node before we turn to the quantum interference involving their respective photons.

\subsection{Two quantum nodes}
The first node is based on the DLCZ protocol, producing heralded photons.
It has the capability to generate entangled photon pairs with an in-built quantum memory that can be used for synchronization of the network.
The DLCZ protocol starts by sending a train of write pulses to ensemble 1, detuned by ${\Delta = \SI{-40}{\mega\hertz}}$ from the ${\ket{g_{1}} = \ket{5S_{1/2}, F=2, m_F=+2}}$ to ${\ket{e_1} = \ket{5P_{3/2}, F=2, m_F=+1}}$ transition.
With a low probability, this generates a write photon (Stokes photon) in the heralding mode which, upon detection at detector SPAD1, heralds a collective spin excitation in the spin state $\ket{s} = \ket{5S_{1/2}, F=1, m_F=0}$.
After a programmable time that is set to synchronize the photon generation from two nodes ($\SI{2.8}{\micro\second}$), we then send a read pulse resonant with the $\ket{s} \rightarrow \ket{e_{1}}$ transition that maps the collective excitation into a read photon (anti-Stokes photon).
This photon is resonant with the $\ket{g} \rightarrow \ket{e_{1}}$ transition and is emitted into a well-defined mode that depends on the phase matching conditions of the process.
For the DLCZ source to generate photons efficiently, the optical depth ($\mathrm{OD}$) of the ensemble must be large \cite{Simon2007}\footnote{Nevertheless, OD shouldn't be too large since it can lead to incoherent absorption and reduction of efficiency.} and the lifetime of the collective excitation must be longer than the synchronization time between the write and the read pulse.
To fulfill both requirements, we employ conventional methods of magneto-optical trapping (MOT) aided by a single retro-reflected beam dipole trap at $\SI{797}{\nano\meter}$.
This dipole trap is useful to increase the duty-cycle of the experiment \footnote{However, the increase in duty cycle is partial, since the writing process of the DLCZ protocol heats the atoms and is associated with rapid loss of optical depth. Still, when producing photons with low $g^{(2)}_{n1}(0)$, we were able to keep a sufficient optical depth during the $\SI{4}{\milli\second}$ of interrogation time.}.
We obtain a cloud with $\mathrm{OD} = 6$ and an initial temperature of $\sim\SI{80}{\micro\kelvin}$.
Furthermore, we use Fabry-Perot cavities (FPC) to filter out unwanted background noise that comes from the write and read pulses, increasing both the efficiency of the heralding and the purity of the generated photons.
The probability to generate a read photon conditionned on the detection of a write photon is around 25\% with these experimental parameters.
Finally, an acousto-optical modulator is employed to adjust the frequency of the read photons and to adjust the photon flux, aligning it with that of the other node.

\begin{figure}[t]
    \begin{minipage}{0.69\linewidth}
        \centering
        \includegraphics[width=\textwidth]{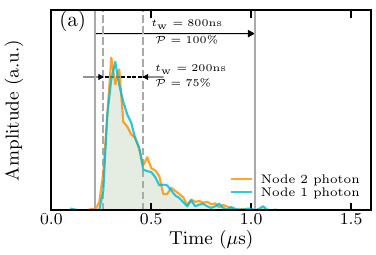}
    \end{minipage}
    \begin{minipage}{0.29\linewidth}
        \centering
        \includegraphics[width=\textwidth]{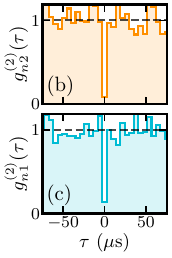}
    \end{minipage}
    \caption{Quantum nodes properties. 
    (a): Temporal modes of the photons emitted from Node 1 and Node 2.
    The gray straight lines represent the temporal windows in which the coincidences of the two-photon interference experiment are counted.
    (b) and (c): Single-photon autocorrelation of the photons emitted from Node 1 an Node 2.
    We measure ${g^{(2)}_{n1}(0)= \num{0.13 \pm 0.04}}$ for the photon issued of Node 1 and  ${g^{(2)}_{n2}(0)= \num{0.09 \pm 0.03}}$ for the photon coming out of Node 2.}
    \label{fig2}
\end{figure}

The second node is based on a blockaded Rydberg ensemble generating single photons in a deterministic fashion. 
Thanks to a two-beam crossed dipole trap at \SI{852}{\nano\meter}, we create a small spherical atomic cloud with size of $\SI{13.5}{\micro\meter}$ (FWHM), $\mathrm{OD} = 12$ and temperature of $\sim\SI{40}{\micro\kelvin}$.
We send a laser pulse resonant with the ${\ket{g_2} = \ket{5S_{1/2}, F=2} \rightarrow \ket{e_2} = \ket{5P_{3/2}, F=3}}$ transition together with a counter-propagating coupling beam resonant with the ${\ket{e_2} \rightarrow \ket{r} = \ket{103S_{1/2}}}$ transition, effectively establishing Rydberg electromagnetically-induced transparency that creates a propagating Rydberg polariton in the cloud \cite{Pritchard2010,Petrosyan2011}.
The Rydberg blockade radius is $r_{b} \approx \SI{16}{\micro\meter}$ at the $n=103$ Rydberg state \cite{Weber2017}.
Since it is larger than the cloud size, the strong dipole-dipole interactions prevents the existence of multiple Rydberg excitations in the cloud \cite{Lukin2001}, allowing us to retrieve a single photon at the output of the cloud with a generation probability of 14\%. 
During the Rydberg polariton's propagation, we switch off-and-on the coupling field (a process known as Dynamical EIT), that effectively freezes the propagation and stores the polariton in the Rydberg state, what is known to enhance the non-linearity and the single photon's quality \cite{Distante2016,Padron-Brito2021a}.

Due to the different nature of the two quantum nodes, their respective experimental cycles differ from one another.
During the $\SI{200}{\milli\second}$ of interrogation time of Node 2, Node 1 performs 17 full cycles, each one containing $\SI{4}{\milli\second}$ of interrogation time.
During each cycle, Node 1 tries to herald a spin wave each $\SI{4}{\micro\second}$.
Once a write photon is detected on SPAD1, an electronic signal is sent to Node 2 which then generates a photon that either temporally overlaps with Node 1 photon (indistinguishable case), or is separated in time (distinguishable case). 
Every half an hour, the experiment is automatically put in calibration mode for around ten minutes in order to compensate for any possible slow drift.
A detailed description of our experimental cycle and the communication between the two nodes is given in the appendix.
In the distinguishable case, the final triple-coincidence rate is around $0.02$ events per second.

\subsection{Photon mode-matching}
Both photons are sent to a measurement platform consisting of a fiber beam-splitter (BS) and two single-photon detectors SPAD2a(b). 
If the two photons are completely indistinguishable, they bunch together at the output of the beam-splitter and no coincidence between SPAD2a and SPAD2b is recorded. 
Truly indistinguishable photons can be obtained only if all their degrees of freedom (spatial, polarization, frequency, time) are mode-matched with one another. 
The spatial mode matching can be ensured by using a single-mode fiber beam-splitter.
Since the polarization of the incoming photons needs to be well controlled at the position of the BS, we use polarization-maintaining fibers and a set of half-wave and quarter-wave plates before the fiber incoupling. 
Injecting classical light at slightly different frequencies into the setup and adjusting the beating fringes, we then measure a polarization-matching greater than \SI{99.9}{\percent}.
The frequency offset (\SI{266}{\mega \hertz}) determined by the atomic levels used to generate the photons in each quantum node is compensated by sending the heralded photons from the DLCZ node to a single-pass acousto-optical modulator (AOM). 
The temporal offset can be controlled by changing the synchronizing time (or storage time) of each quantum node.
Finally, assuming that both photons are Fourier limited, their frequency-mode is completely determined by their temporal mode.
We match the temporal mode of the DLCZ produced photon to that of the Rydberg-produced photon by modulating the intensity of the read pulse in the DLCZ protocol \cite{Farrera2016}, allowing us to maintain a temporal mode-matching higher than $98\%$ as represented in \autoref{fig2}(a).
In both cases, we produce a photon with a steep leading edge of \SI{60}{\nano\s} and resembling an exponential decay trailing edge with $1/e$ time of around $\SI{180}{\nano\second}$. 
The photon flux is also adjusted by changing the driving amplitude of the AOM, such that the probability to have a photon at the beamsplitter is similar for each source (around 2.8\% per trial).

To evaluate the quality of each photon, we measure the autocorrelation function $g^{(2)}_{n1}(\tau=0)$ and $g^{(2)}_{n2}(\tau=0)$. 
The results are displayed in \autoref{fig2}(b), where the counts are integrated over a temporal window of length $t_{\mathrm{w}} = \SI{800}{\nano\second}$ in each \SI{5.4}{\micro\second}-long bin, representing one experimental trial.
We measure $g^{(2)}_{n1}(0)= \num{0.13 \pm 0.04}$ for the photon generated by node 1 and  $g^{(2)}_{n2}(0)= \num{0.09 \pm 0.03}$ for the photon coming from node 2.


\begin{figure}[t]
    \begin{minipage}{\linewidth}
        \centering
        \includegraphics[width=0.95\textwidth]{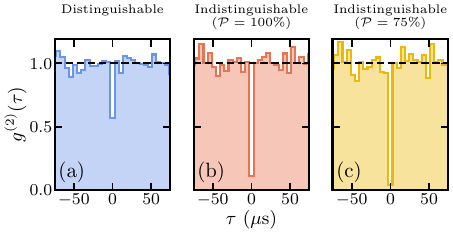}
    \end{minipage}
    \begin{minipage}{\linewidth}
        \centering
        \includegraphics[width=0.99\textwidth]{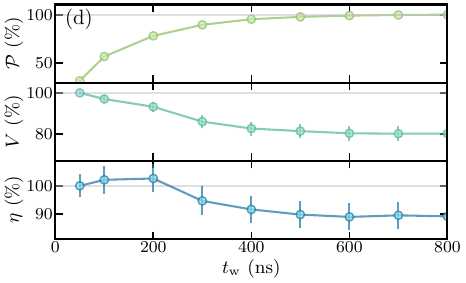}
    \end{minipage}
    \caption{Two-photon interference.
    (a), (b), (c): Normalized coincidence rate $g^{(2)}(\tau)$ between SPAD2a and SPAD2b in the distinguishable case (blue) and indistinguishable case for the whole pulse (red), and for a proportion $\mathcal{P}=75\%$ of counts (yellow).
    (d): Measured visibility ($V$), indistinguishability ($\eta$) and photon counts share ($\mathcal{P}$), for various lengths of the temporal window. } 
    \label{fig:results}
\end{figure}
\section{Two-photon interference}
We now move to the two-photon interference between the two quantum nodes.
In both indistinguishable case and distinguishable case, we record coincidences to measure the visibility of the two-photon interference which is given by
\begin{equation}
    V = 1 - \frac{g^{(2)}_{i}(0)}{g^{(2)}_{d}(0)}\text{,}
\end{equation}
where $g^{(2)}_{d}(0) (g^{(2)}_{i}(0))$ is the normalized coincidence rate between SPAD2a and SPAD2b in the distinguishable (indistinguishable) case.
Coincidences are recorded in a specific temporal window as illustrated on \autoref{fig2} where a window of length ${t_{\mathrm{w}}=\SI{800}{\nano\second}}$ contains ${\mathcal{P}=\SI{100}{\percent}}$ of photon counts and another window of length ${t_{\mathrm{w}}=\SI{200}{\nano\second}}$ contains a proportion of ${\mathcal{P}=\SI{75}{\percent}}$. 

\begin{figure}
    \centering
    \includegraphics[width=0.48\textwidth]{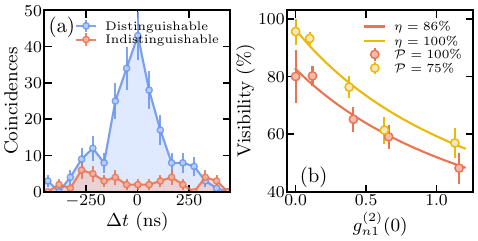}  
    \caption{(a): Number of detected coincidences separated by a time delay $\Delta t$, within the analysis window $t_{\mathrm{w}}=\SI{800}{\nano\second}$.
    (b): Measured two-photon visibilities for different values of DLCZ photons autocorrelation $g^{(2)}_{n1}$, for the temporal window containing $\mathcal{P}= \SI{100}{\percent}$ and \SI{75}{\percent} of the counts.
    The solid lines are the expected values estimated from \autoref{eq2} for an indistinguishability of $\eta = \SI{86}{\percent}$ (red) and $\eta = \SI{100}{\percent}$ (yellow). }
    \label{fig:time_resolved}
\end{figure}

For counts integrated over the full window (${t_{\mathrm{w}}=\SI{800}{\nano\second}}$), we measure a visibility of ${V=\SI{80.2 \pm 3.6}{\percent}}$ with ${g^{(2)}_{d}(0)=\num{0.57 \pm 0.05}}$ and ${g^{(2)}_{i}(0)=\num{0.11 \pm 0.02}}$, as shown in \autoref{fig:results}(a) and \autoref{fig:results}(b).
This result is, to our knowledge, the highest reported visibility of a two-photon interference in hybrid systems, and it is higher than most realizations with nodes made of the same platform without discarding photon counts.
If we take ${\mathcal{P}=\SI{75}{\percent}}$ of all photon counts (${t_{\mathrm{w}} = \SI{200}{\nano\second}}$), we measure a visibility as high as ${V=\SI{93.2 \pm 2.3}{\percent}}$, with ${g^{(2)}_{i}(0)=\num{0.04\pm0.01}}$.
We further confirm the strong suppression of the coincidence rate for overlapping pulses by looking at the time-resolved HOM interference, as shown in \autoref{fig:time_resolved}(left).
When the photons are set to be distinguishable, we observe a peak of coincidences (blue data) around zero delay that is linked to the convolution of the photons' temporal mode.
One can appreciate a strong suppression of the coincidence number (red data) over the whole temporal window.
We emphasize here that the results presented in this paragraph were measured with raw coincidences and without correction of any kind.

To estimate the indistinguishability $\eta$ between the two photons, we take into account the fact that the non-zero value of the single-photon autocorrelation $g^{(2)}(0)$ (see \autoref{fig2}) reduces the visibility of the interference.
For a 50/50 beamsplitter, assuming that each node emits at most two photons, the indistinguishability $\eta$ can then be evaluated using the following formula \cite{Craddock2019}:
\begin{equation}\label{eq2}
V = \frac{2\eta}{g^{(2)}_{n1}(0)x+g^{(2)}_{n2}(0)/x+2}
\end{equation}
where $g^{(2)}_{n1}(0) (g^{(2)}_{n2}(0))$ is the autocorrelation of the photon incoming from Node 1(2) and $x=p_{n1}/p_{n2}$ is the ratio between the probabilities to detect a photon from Node 1 and 2.
We infer an indistinguishability ${\eta = \SI{89.2 \pm 4.9}{\percent}}$ for ${\mathcal{P} = \SI{100}{\percent}}$. 
The indistinguishability increases to \SI{94.6 \pm 5.2}{\percent} for ${\mathcal{P} = \SI{90}{\percent}}$ and becomes compatible with unity within error bars for $\mathcal{P} = \SI{75}{\percent}$, showing that both photons exhibit a near-perfect indistinguishability between one another.
To further study this interference, we repeat this measurement for different values of $g^{(2)}_{n1}(0)$ to see if the scaling of the visibility in \autoref{eq2} is well respected.
$g^{(2)}_{n1}(0)$ can be changed by varying the write pulse power in the writing stage of the DLCZ protocol, which changes the probability to generate more than one atomic excitation.
These measurements are plotted in \autoref{fig:time_resolved} alongside the predictions of \autoref{eq2} (solid lines) with parameters ${x=1.3, \  g^{(2)}_{n2}(0)=0.1}$ and where the indistinguishability is $\eta = \{ 86, 100\}\%$ for $\mathcal{P} = \{ 100, 75\}\%$.
We observe a good agreement of our data with the formula, indicating that our visibility is indeed reduced due to non-zero two-photon suppression of each photon.

Finally, we plot in \autoref{fig:results}(d) the visibility $V$, indistinguishability $\eta$ and proportion $\mathcal{P}$ of photon detections as a function of the length of the temporal window $t_{\text{w}}$.
In the limit of large temporal window, we observe a saturation of indistinguishability $\eta$ around $90\%$, meaning that the photons' mode-matching is not perfect.
We attribute this to small spectral impurity, meaning that our photons might not be completely Fourier limited.
By decreasing $t_{\text{w}}$, we effectively broaden our frequency mode, allowing for a better overlap in frequency space, and thus higher indistinguishabilities.
For completeness, we explain here why some values of $\eta$ were found to be above unity for small $t_{\text{w}}$. 
The estimator of \autoref{eq2} is bounded, in the sense that if all the relevant quantities ($g^{(2)}_{\text{n1}}(0)$, $g^{(2)}_{\text{n2}}(0)$, $V$ and $x$)  were measured simultaneously, it would not be possible to obtain a value of indistinguishability larger than unity. 
However, in our experiment we had to measure these values separately and we were therefore exposed to statistical fluctuations. 
If the indistinguishability is close to unity, the probability to find $\eta > 100\%$ is therefore high.
We emphasize here that our measurements are compatible with unity within error bars, as the error bar goes down to 97\%. 

Throughout the experiment, the input probe mean photon number of Node 2 was kept low to maintain low $g^{(2)}_{n2}(0)$.
However, it is important to note that a high mean input photon number would not just degrade the $g^{(2)}_{n2}(0)$, but it could also diminish the purity of the single photons generated by Node 2, as predicted by Gorshkov et al. [50].
This decrease in purity would consequently lead to a reduction in the indistinguishability of the photons.
We report on the observation of this effect and analyse our results in the appendix \autoref{appendix:Gorshkov} of this manuscript.

\section{Discussion and outlook}
In summary, our work demonstrates high-visibility interference of two photons produced in dissimilar and independent quantum nodes. 
We achieve high indistinguishability by carefully mode-matching the two photons in all degrees of freedom, without the need of discarding photon detections.
If this setup was used for a quantum communication protocol, such as generating remote entanglement between Node 1 and Node 2, the maximum achievable fidelity of the entanglement with respect to the ideal Bell-state would be equal to $F = (1-\eta)/2$ \cite{Craddock2019} for ideal single photons.
Based on the data presented in Fig 4.(b), it means that without discarding any photon detection, we would be limited to $F = 93 \%$. 
By keeping a proportion $\mathcal{P}=75\%$ of the counts, thus reducing the efficiency of the protocol by only $25\%$, a fidelity close to $F = 100 \%$ would be theoretically achievable.
For such an experiment, one would need to generate entangled photon pairs in Node 2, as done for example in \cite{Sun2022b}, and perform a Bell-state measurement \cite{Sangouard2011}.
Our setup in Node~2 would need to be upgraded to gain the ability of addressing different Rydberg states at the same time and coherently control these single Rydberg excitations with high fidelity \cite{Sun2022b}.
In this context, this experiment opens the path towards remote entanglement of hybrid systems, and connection of quantum processors through quantum repeater links.

\appendix
\begin{figure*}[]
    \begin{minipage}[c]{0.7\linewidth}
   \def\svgwidth{\columnwidth}
   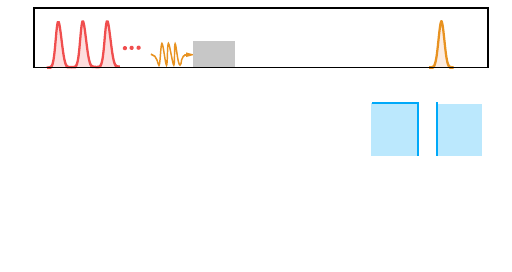
    \end{minipage}
    \begin{minipage}{0.28\linewidth}
        \centering
        \includegraphics[width=\textwidth]{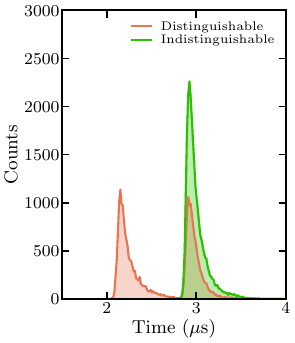}
    \end{minipage}
    \centering
    \caption{Distinguishable and indistinguishable cases. 
    (left): Schematics of one experimental trial.
    Quantum Node 2 keeps trapping atoms until it receives a classical signal from Node 1, indicating that it has heralded a write photon, and will produce a read photon after \SI{2.8}{\micro\second}.
    After a programmable synchronization time, Node 2 produces a photon using Rydberg EIT.
    Depending on this time, the photons emitted from Node 1 and Node 2 either overlap temporally (indistinguishable case) or don't (distinguishable case).
    (right): Typical histogram obtained in the distinguishable case (red) and indistinguishable case (green).}
    \label{fig:sequence}
\end{figure*}

\section{Node properties}
\subsection{Node 1}
In quantum node 1, $^{87}$Rubidium atoms are trapped from background-gas pressure in a magneto-optical trap (MOT) during \SI{8}{\milli\second}.
Subsequently, the cloud is cooled by polarization-gradient cooling (PG cooling) for \SI{500}{\micro\second}, while ramping down the magnetic field.
Atoms are then prepared in the ground state ${\ket{g_1}}$ by Zeeman optical pumping, achieving an initial optical depth of $\mathrm{OD} = 6$ on the ${\ket{g_1}} \rightarrow \ket{e_1}$ transition.
To reduce atom loss by diffusion and fall in gravity, the atoms are loaded into an optical dipole trap which intersects with the ensemble.
The dipole trap consists of a focused light beam of linear polarization, at \SI{796.5}{\nano\meter}.
\SI{250}{\milli\watt} of power are focused down to a beam waist of \SI{110}{\micro\meter}.
The light is retro-reflected, forming a one-dimensional lattice with potential depth of $U_0 \approx \SI{400}{\micro\kelvin}$, incident at an angle of $\sim \ang{10}$ with the photon mode.
Only then starts the interrogation time, where the write photon generation starts by sending a train of write pulses.
An unsuccessful attempt is followed by a cleaning pulse, pumping the atoms back to $\ket{g_1}$.
A successful heralding is instead followed by an electronic trigger sent to quantum node 2, signaling the successful generation of a spin wave.
\SI{2.8}{\micro\second} after the write pulse, a read-out pulse retrieves the excitation as a photon.
The $1/e^2$ beam radius is \SI{69}{\micro\meter} for the photon mode and \SI{180}{\micro\meter} for the excitation pulses.
After \SI{4}{\milli\second} of interrogation time, the magnetic field is switched on again and the sequence repeats.
The total cycle lasts $\sim \SI{12}{\milli\second}$.

The frequency of the photon is, however, not compatible with the photon produced in Node 2, so it is shifted by +\SI{266}{\mega\hertz} with an acousto-optic modulator to make it resonant to the $\ket{g_2} \rightarrow \ket{e_2}$ transition.

\begin{figure*}
    \centering
    \includegraphics[width=\textwidth]{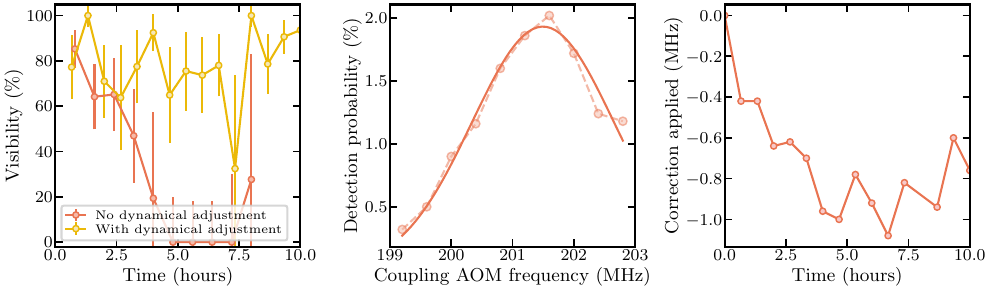}
    \caption{Frequency stabilization in Node 2.
    (left): Two-photon interference visibility as a function of measurement time.
    The experiment drifts after a few hours (red data).
    To counteract these drifts, we dynamically adjust the coupling beam frequency as described in the text.
    The stability is greatly improved, as represented in yellow (corresponding to the data presented in the main text).
    (middle): By scanning the coupling frequency, we are able to find the Rydberg EIT resonance.
    This measurement is performed every half hours, and the frequency is readjusted accordingly.
    The solid line is a Gaussian fit to the data.
    (right): Example of the applied correction on the frequency of the coupling laser to cancel for the resonance drift.
    In this case, the correction corresponds to the yellow data of the left plot.}
    \label{fig:Freq_stabilization}
\end{figure*}

\subsection{Node 2}

In quantum node 2, $^{87}$Rubidium atoms are loaded in a crossed dipole trap from a MOT.
We load a MOT during \SI{1}{\second}, followed by a temporal dark-spot compression to increase the atomic density, in which the magnetic field gradient is ramped up from 20 G/cm to 50 G/cm during $\SI{9}{\milli\second}$.
The atoms are then cooled with polarization-gradient cooling to $\approx \SI{40}{\micro\kelvin}$ for \SI{8}{\milli\second} and prepared in the ground state $\ket{g_2}$.
The dipole trap consists of two tightly focused light beams of linear polarization at \SI{852}{\nano\meter}, each with a power of \SI{1}{\watt} and a beam waist of \SI{34}{\micro\meter}.
One beam is horizontal, in-plane with the coupling beam and photon modes, and intersects with the photon mode at an angle of \ang{22}.
The second beam is incident from the top.
This way, a spherical cloud with $\mathrm{OD} = 12$ on the $\ket{g_2} \rightarrow \ket{e_2}$ transition is obtained, with a diameter of \SI{13}{\micro\meter} (FWHM).
The EIT coupling field is counter-propagating and resonant with the ${\ket{e_2} \rightarrow \ket{r}}$ transition, corresponding to a wavelength of \SI{479.3118}{\nano\meter}, derived from a 479-nm frequency converted amplified diode laser with a seed at \SI{958}{\nm}.
The $1/e^2$ beam radius is \SI{6.5}{\micro\meter} for the probe and \SI{13.5}{\micro\meter} for the coupling mode.
The total interrogation time is \SI{200}{\milli\second}, in a cycle that lasts $\sim \SI{1.25}{\second}$.

Depending on the measurement, we time the probe and coupling field with respect to the classical signal sent by Node 1 to either overlap the generated photon with the Node 1 photon, or separate it in time.
We call these two cases ``indistinguishable" and ``distinguishable", as presented in \autoref{fig:sequence}.

\section{Frequency stabilization}

Reducing the linewidths and maintaining good stability of all lasers frequencies used in each quantum node is important in our experiment. 
This is because the visibility of the two-photon interference is sensitive to the frequency offset between each photon, which can be affected by various frequency drifts.
In this paragraph, we start by describing our locking systems in Node 1 and in Node 2. We then discuss several instability problems that appeared in Node 2, and we finally explain the method we implemented to fix them, in order to acquire a good stability over several hours of measurements.
Data for the stabilization improvement is presented in Fig \ref{fig:Freq_stabilization}.

In Node 2, we use a cascaded lock of two main elements: modulation transfer spectroscopy (MTS) \cite{Camy1982} and a transfer cavity.
We lock our probe laser at \SI{780}{\nano\meter} to the MTS signal using a method described in \cite{Escobar2015}.
Careful alignment of the MTS setup allowed us to reduce frequency drifts to below \SI{20}{\kilo\hertz} (limited by residual amplitude modulation introduced by an electro-optic modulator).
A fraction of the same laser light is sent to a \SI{25}{\cm}-long plano-concave transfer cavity of finesse $\sim 1800$ whose length is controlled by a mirror fixed to a piezo actuator.
The cavity length is stabilized with the standard Pound-Drever-Hall (PDH) technique \cite{Drever1983}.
A small portion of the coupling laser seed light at \SI{958}{\nm} is sent to the same transfer cavity and its frequency is also locked with PDH technique.
This allows for not only stabilization of the central frequency, but also to narrow the laser linewidth to $\sim \SI{60}{\kilo\hertz}$, as estimated by comparing the root-mean-square of the error signal and its slope. 
In Node 1, one laser at \SI{780}{\nano\meter} is locked to a saturated absorption spectroscopy signal offering stability on the order of \SI{200}{\kilo\hertz}, used for the writing pulses and magneto-optic trapping of the atoms. 
Another laser at 780nm, used for reading-out the collective excitation is locked with a phase-lock loop (PLL) technique to the probe laser of Node 2, reducing uncorrelated drifts between the two nodes. 

In the context of high visibility of photon-photon interference, we were affected by a slow drift of the Rydberg resonance and of the coupling laser frequency in Node 2, attributed to several reasons.
When locked for several hours, the coupling laser residual amplitude modulation causes a slow frequency drift, effectively changing the frequency of Node 2 photons.
Additionally, Node 2 photon's frequency was affected by a shift of the Rydberg EIT resonance over the course of hours, possibly due to slow changes of the ambient electric field. 
The level $103S_{1/2}$ is indeed highly sensitive to electric field, and we estimate (using ARC package \cite{Sibalic2017}) that a change of electric field strength as small as \SI{10}{\milli\volt\per\centi\m} is enough to shift the energy level by $\sim\SI{1}{\mega\hertz}$.
Last, we observed that the trigger rate of the Rydberg node affects the value of the EIT resonance, if the rate is too low.
This means that the resonant frequency depends on the rate of write photon detections in Node 1.
We do not have a clear explanation why this happens, but we understand that it is linked to the total time in which the coupling laser is shined onto the atoms, and we suspect a technical issue. 

Since our experiment required to perform measurements for several tens of hours, we implemented fixes in the experimental sequence to keep the resonance stable.
Regarding the trigger rate issue, we understood that the resonance was affected by the amount of time the blue power was turned on during the interrogation time.
To overcome this effect, we decided to turn on the coupling laser for $\SI{10}{\milli\second}$ before the start of the interrogation time, allowing us to go out of the range at which the trigger rate influences the resonance.
This fixed most of the visibility decay with measurement time.
Then, to correct for both frequency drift of the coupling laser, and EIT resonance drift, we implemented a dynamical method that probes the atomic resonance and re-adjust the laser every half an hour to keep the coupling laser on two-photon resonance.
This method consists in storing weak pulses of probe light in the Rydberg ensemble using Rydberg EIT.
We then scan the coupling light frequency and find the resonance by finding the maximum of storage efficiency. 
The Hong-Ou-Mandel experiment between the two nodes then starts again with the corrected coupling frequency, until another calibration is done half an hour later.
The results for such a calibration sequence are presented on \autoref{fig:Freq_stabilization}.

\section{Observation of dissipative interactions in the Rydberg Node}
\label{appendix:Gorshkov}
\begin{figure*}
    \centering
    \includegraphics[width=\textwidth]{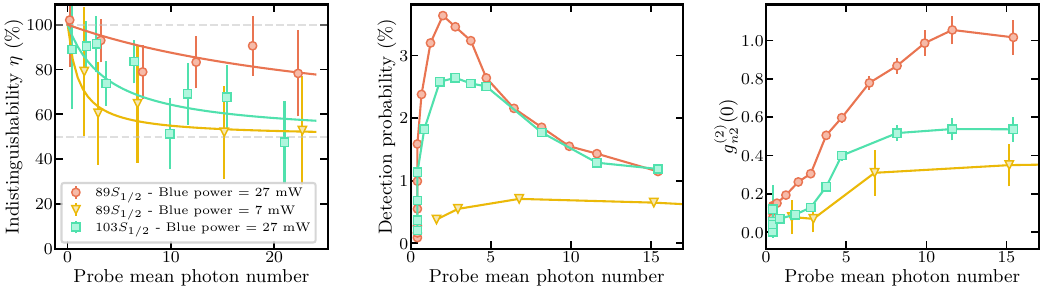}
    \caption{Dissipative interactions in the Rydberg node.
    (left): Indistinguishability $\eta$ of the two-photon interference for different input probe mean photon number in Node 2, measured for different conditions of Rydberg blockade radius by changing the Rydberg level or the EIT linewidth.
    We observe a decrease of $\eta$ for increasing probe mean photon number, a sign of reduced purity of the generated photon due to dissipative interactions in the Rydberg medium.
    The solid lines are a guide to the eye, based on the purity formula proposed in \cite{Gorshkov2013}, where we weighed the mean input photon number of each curve differently.
    The weights are free fitting parameters, and we did not find any meaningful relation between the weights and the blockade radii.
    (middle): Rydberg photon detection probability as a function of input mean photon number.
    Above a few photons per pulse, the generation probability decreases, possibly due to the EIT filtering effect \cite{Zeuthen2017} and pollutants \cite{Bienias2020}.
    (right): Variation of the antibunching $g^{(2)}_{n2}(0)$ of the generated photon with respect to input probe mean photon number.
    For higher blockade radius, we obtain lower values of $g^{(2)}_{n2}(0)$.
    The data presented in Fig. \ref{fig:Gorshkov} (left) was taken with a lower optical depth and lower maximum available coupling power than the data taken for \autoref{fig:Gorshkov} (middle and right).
    This is because these measurements were performed at different times, when experimental conditions were different.}
    \label{fig:Gorshkov}
\end{figure*}
Theoretical predictions by Gorshkov and collaborators \cite{Gorshkov2013} suggested that photons generated under Rydberg EIT (rEIT) conditions with a high mean number of input probe photons would be in a mixed state.
It results from the fact that blockaded ensembles allow only one Rydberg excitation at a time.
This means that under perfect rEIT conditions, only the initial photon propagates as a polariton, while subsequent ones scatter away.
These scattering events occur within a distance approximately equal to the blockade radius, defined as the point where the interaction strength matches the single-atom EIT linewidth, and is typically given by \cite{Firstenberg2016}
\begin{equation}
    \label{eq:blockade_radius}
    r_{b} = \sqrt[6]{\frac{C_{6} \Gamma}{\Omega_{c}^2}},
\end{equation}
where $C_{6}$ is the van der Waals coefficient (controlling interaction strength), $\Omega_{c}$ is the coupling Rabi frequency, and $\Gamma$ is the decay rate of the excited state.
Both, the van der Waals coefficient and the coupling Rabi frequency scale with $n$, i.e. $C_6 \propto n^{11}$ and $\Omega_c \propto n^{-3/2}$.

As photons scatter at positions determined by the polariton, they carry information about its location.
Therefore, each scattering event decreases the purity of the polaritonic state and, consequently, of the retrieved photonic state.
Put differently, the scattered photons become entangled with the polariton, and tracing them out from the description leaves the polariton in a mixed state.
The expected outcome is a reduction in the photon's purity with an increase in the mean number of input photons.
Moreover, this phenomenon is associated with the spatial localization of the Rydberg polariton, where successive scattering events pinpoint the polariton's position within the cloud.
This localization in space broadens the polariton in frequency, potentially causing additional absorption if it surpasses the EIT linewidth \cite{Zeuthen2017}.
Consequently, we anticipate a decreased generation probability for larger input mean photon numbers.

We can study the reduction of the purity of the Rydberg photons owing to the fact that the visibility of the HOM interference depends on the purity of the input photon states.
For perfectly separable single photons with identical spectral amplitudes \footnote{
Which is a reasonable assumption in our case, as both of the photons are expected to be Fourier limited and their temporal modes are well-matched.
}, the HOM visibility is expressed as ${V = \mathrm{Tr}(\rho_{n1}\rho_{n2})}$ \cite{Branczyk2017}.
Here, ${\rho_{n1}}$ and ${\rho_{n2}}$ refer to the density matrices of the interfering photons, while ${\mathrm{Tr}}$ signifies the matrix trace.
As our photons are not perfect and exhibit a $g^{(2)}(0)$ above zero, one can equate this visibility with the indistinguishability $\eta$ discussed in the main text.
Since we observe $\eta$ to be close to one, it suggests that ${\rho_{n1} \approx \rho_{n2}}$.
Consequently, any alteration of ${\rho_{n2}}$ will reduce the indistinguishability, specifically a reduction of its purity ${\mathrm{Tr}(\rho_{n2}^2)}$.

The measurements revealing purity reduction of the Rydberg photons, as described above, are presented in \autoref{fig:Gorshkov}. 
We observe a general trend of decreasing indistinguishability $\eta$ for increasing input probe mean photon number. 
When we change the Rydberg level from $103S_{1/2}$ (turquoise squares) to $89S_{1/2}$ (red circles) keeping the same coupling power \footnote{
Lowering the Rydberg state affects both the interaction strength and the coupling Rabi frequency, however, the net effect is the reduction of the blockade radius.
}, consequently reducing the Rydberg blockade radius ${r_{b}}$ from \SI{15.5}{\micro\meter} to \SI{11}{\micro\meter} which is less than the ensemble size, we observe a significantly weaker decay of the indistinguishability with input probe mean photon number.
This is an expected behavior --- for a fixed input probe mean photon number, if the Rydberg blockade radius is decreased below the ensemble size, the number of scattering events is reduced and the purity of the photonic state is higher (at the expense of increased $g^{(2)}(0)$).
We then increase again the blockade radius to \SI{13}{\micro\meter} by reducing the coupling Rabi frequency at $89S_{1/2}$ (yellow triangles), and observe a stronger decay of $\eta$.
We do not know the reason why this decay is stronger than in the \SI{15.5}{\micro\meter} case.

\begin{figure}
    \centering
    \includegraphics[]{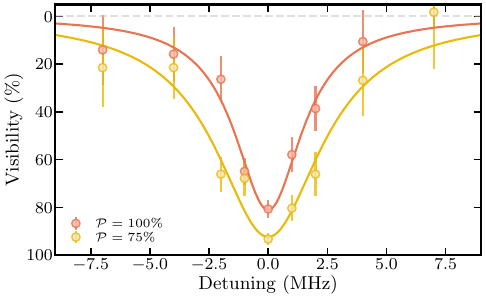}
    \caption{Hong-Ou-Mandel dip in frequency space.
    The frequency of the AOM is varied to change the detuning between the two photons.
    The FWHM of the dip is larger for the temporal window containing a smaller proportion $\mathcal{P}$ of the counts.
    Solid lines are Lorentzian fit to the data, as expected for exponential photons.}
    \label{fig:Dip}
\end{figure}

These results are, to our knowledge, the first observation of the effects predicted in \cite{Gorshkov2013}.
Nevertheless, notable disparities exist between the theory and our experiment.
Gorshkov et al. assumed the pulses to be fully compressed within the medium, a condition not met in our experiment.
However, by performing the storage we effectively discard parts of the pulse that were not affected by the blockade effect, bringing it closer to the theory conditions.
At the same time, the storage is known to enhance the Rydberg medium nonlinearity \cite{Distante2016, Lowinski2024}, deviating our experiment from the theory consideration.

Moreover, limitations of our experimental data hinder definitive conclusions.
First, our results exhibit large error bars for large input mean photon number and low $\Omega_{c}$, limited by low generation probability as pictured in \autoref{fig:Gorshkov} (middle), possibly due to the EIT filtering effect as described earlier and so-called pollutants \cite{Bienias2020}.
Additionally, even though we kept the same DLCZ excitation probability in Node 1 for this measurement, we had to adapt the read pulse waveform since the photons produced with smaller $\Omega_{c}$ in the Rydberg medium have a longer wave shape. 
This might also have changed Node 1's photons properties for each configuration in this measurement.

\section{Hong-Ou-Mandel dip}

To further explore the properties of our photon-photon interference, we measure a Hong-Ou-Mandel dip in the frequency domain as plotted in \autoref{fig:Dip}.
We scan the frequency difference of the two photons by changing the frequency of the AOM of Node 1.
We observe a dip resembling a Lorentzian, as expected for photons having an exponential temporal waveform.
Depending on the length of the temporal window used for analysis, the FWHM of the dip is modified, a nice example that shows that reducing the temporal-window is equivalent to artificially increasing the bandwidth of the photons, thus increasing the overlap of their modes.

\section{Visibility calculation}
We explain here our choice to evaluate the visibility of the two-photon interference with the normalized coincidence probability in the case where the photons are distinguishable $g^{(2)}_{d}(0)$ compared to the case in which they are indistinguishable $g^{(2)}_{i}(0)$, following
\begin{equation}\label{eq:visi}
    V = 1 - \frac{g^{(2)}_{i}(0)}{g^{(2)}_{d}(0)},
\end{equation}
which is equation (1) in the main text.
We argue that this approach is more conservative and more reliable compared to estimating visibility through direct comparison of raw coincidences in distinguishable and indistinguishable cases, a common practice in many studies. 

For both methods of calculating the visibility, it is necessary to measure independently distinguishable and indistinguishable cases. 
This makes the results susceptible to experimental drifts, such as fluctuations in photon generation efficiency or optical losses over the extended experimental time. 
For instance, if the photon generation rate dropped during indistinguishable coincidence measurements compared to distinguishable ones, the visibility would be erroneously higher. 
By utilizing normalized coincidence probabilities, these type of fluctuations are effectively mitigated.
It is important to note that while this method enhances reliability, it doesn't eliminate the inherent statistical uncertainties of our measurements, and the effect of other type of experimental drift which might also affect the result.

In the following calculation, we explain how we obtain \autoref{eq2} of the main text.
We start by estimating the probability $\mathcal{P}$ of having one photon at each output port of the beamsplitter (BS) per experimental trial.
For a 50/50 BS :
\begin{equation}\label{eq:P}
    \mathcal{P} = p_{1}^{(n1)} p_{1}^{(n2)} (1 - P_\mathrm{HOM}) + \frac{1}{2} p_{2}^{(n1)} + \frac{1}{2} p_{2}^{(n2)}
\end{equation}
where $p_{1}^{(n1)}$ ($p_{1}^{(n2)}$) is the probability that a single photon from Node 1 (Node 2) arrives at the BS, $p_{2}^{(n1)}$ ($p_{2}^{(n2)}$) is the probability that two photons from Node 1 (Node 2) arrive at the BS and $P_\mathrm{HOM}$ is the probability that the single photons emitted from Node 1 and Node 2 bunch when they arrive at the BS. The terms containing $p_{2}^{(n1)}$ and $p_{2}^{(n2)}$ are multiplied by 1/2, since the probability that two photons impinging on the BS exit with one photon at each port is equal to 1/2.
We neglect higher-order processes.
It can be shown (for example, in supplementary material of \cite{Craddock2019}) that 
\begin{equation}
    P_\mathrm{HOM} = \frac{1 + \eta}{2}\text{,}
\end{equation}
where $\eta$ is the mode overlap (the indistinguishability) of the photons.
Factoring $p_{1}^{(n1)} p_{1}^{(n2)}$ out of \autoref{eq:P}, we get
\begin{equation}
\begin{aligned}
    \mathcal{P} = {} & 
    p_{1}^{(n1)} p_{1}^{(n2)} \\ &
    \times \left[ 
    \frac{1 - \eta}{2} +
    \frac{1}{2}\frac{p_{2}^{(n1)}}{p_{1}^{(n1)} p_{1}^{(n2)}} +
    \frac{1}{2}\frac{p_{2}^{(n2)}}{p_{1}^{(n1)} p_{1}^{(n2)}}
    \right] \text{.}
\end{aligned}
\end{equation}
We identify that 
\begin{equation}
    g^{(2)}_{n1}(0) \approx \frac{2 p_{2}^{(n1)}}{p_{1}^{(n1)} p_{1}^{(n1)}}
    \quad \text{and} \quad
    g^{(2)}_{n2}(0) \approx \frac{2 p_{2}^{(n2)}}{p_{1}^{(n2)} p_{1}^{(n2)}}\text{,}
\end{equation}
and using the notation $x = p_{1}^{(n1)} / p_{1}^{(n2)}$, we arrive at
\begin{equation}
    \mathcal{P} = p_{1}^{(n1)} p_{1}^{(n2)} \left[ 
    \frac{1 - \eta}{2} + \frac{1}{4}g^{(2)}_{n1}(0)x + \frac{1}{4} g^{(2)}_{n2}(0) \frac{1}{x}
    \right]\text{.}
\end{equation}
We can now express the normalized detected coincidence probabilities as the ratio of having a coincidence click divided by the product of the probabilities of having a single click in detector 2a and detector 2b.
When photons are completely distinguishable, $\eta = 0$ and we have
\begin{equation}
\begin{aligned}
    g^{(2)}_{d}(0) & = \frac{\mathcal{P}|_{\eta=0} \, \alpha_{2a} \alpha_{2b}}
    {\left[ \left( 
    \frac{1}{2} p_{1}^{(n1)} + \frac{1}{2} p_{1}^{(n2)} \right) \alpha_{2a} \cdot \left(
    \frac{1}{2} p_{1}^{(n1)} + \frac{1}{2} p_{1}^{(n2)} \right) \alpha_{2b}
    \right]} \\
    & = \frac{4 x}{\left( 1 + x \right)^{2}}\left[\frac{1}{2} + \frac{1}{4} g^{(2)}_{n1}(0) x + \frac{1}{4} g^{(2)}_{n2}(0) \frac{1}{x}\right], 
\end{aligned}
\end{equation}
where $\alpha_{2a}$ ($\alpha_{2b}$) is the detection efficiency of detector 2a (2b) in Fig 1. of the main text.
Similarly, in the indistinguishable case, we can write
\begin{equation}
    g^{(2)}_{i}(0) = \frac{4 x}{\left( 1 + x \right)^{2}}\left[\frac{1 - \eta}{2} + \frac{1}{4} g^{(2)}_{n1}(0) x + \frac{1}{4} g^{(2)}_{n2}(0) \frac{1}{x} \right] \text{.}
\end{equation}
We note that these quantities are dependent on $p_{1}^{(n1)}$ and $p_{1}^{(n2)}$ only through their ratio $x$ and the autocorrelations $g^{(2)}_{n1}$ and $g^{(2)}_{n2}$.
It means that if there is a global change of detection rate, it does not affect the visibility, which is why this approach is more robust, as explained at the beginning of this section.
Finally, assuming that $x$, $g^{(2)}_{n1}$ and $g^{(2)}_{n2}$ are constant during the both measurement of $g^{(2)}_{d}(0)$ and $g^{(2)}_{i}(0)$, we insert the last two equations into \autoref{eq:visi} to obtain \autoref{eq2} of the main text:
\begin{equation}
    V = \frac{2\eta}{g^{(2)}_{n1}(0)x+g^{(2)}_{n2}(0)/x+2}\text{.}
\end{equation}

\begin{acknowledgments}
This project received funding from the Government of Spain (PID2019-106850RB-I00 project funded by MCIN/ AEI /10.13039/501100011033; Severo Ochoa CEX2019-000910-S), from MCIN with funding from European Union NextGenerationEU (PRTR-C17.I1), from the European Union Horizon 2020 research and innovation program under Grant Agreement No. 899275 (DAALI), from the Gordon and Betty Moore Foundation through Grant No. GBMF7446 to H. d. R, from Fundaci{\'o} Cellex, Fundaci{\'o}  Mir-Puig and from Generalitat de Catalunya (CERCA, AGAUR). 
L.H. and J.L. acknowledge funding from the European Union Horizon 2020 research and innovation program under the Marie Sk\l{}odowska-Curie grant agreement No. 713729.
\end{acknowledgments}

\bibliography{HongOuMandel}

\end{document}

%% file: Setup_HOM_v4.pdf_tex
\begingroup%
  \makeatletter%
  \providecommand\color[2][]{%
    \errmessage{(Inkscape) Color is used for the text in Inkscape, but the package 'color.sty' is not loaded}%
    \renewcommand\color[2][]{}%
  }%
  \providecommand\transparent[1]{%
    \errmessage{(Inkscape) Transparency is used (non-zero) for the text in Inkscape, but the package 'transparent.sty' is not loaded}%
    \renewcommand\transparent[1]{}%
  }%
  \providecommand\rotatebox[2]{#2}%
  \newcommand*\fsize{\dimexpr\f@size pt\relax}%
  \newcommand*\lineheight[1]{\fontsize{\fsize}{#1\fsize}\selectfont}%
  \ifx\svgwidth\undefined%
    \setlength{\unitlength}{385.15812911bp}%
    \ifx\svgscale\undefined%
      \relax%
    \else%
      \setlength{\unitlength}{\unitlength * \real{\svgscale}}%
    \fi%
  \else%
    \setlength{\unitlength}{\svgwidth}%
  \fi%
  \global\let\svgwidth\undefined%
  \global\let\svgscale\undefined%
  \makeatother%
  \begin{picture}(1,0.61087172)%
    \lineheight{1}%
    \setlength\tabcolsep{0pt}%
    \put(0,0){\includegraphics[width=\unitlength,page=1]{Setup_HOM_v4.pdf}}%
    \put(0.97712913,0.55829392){\color[rgb]{0,0,0}\rotatebox{-0.36067414}{\makebox(0,0)[lt]{\lineheight{1.25}\smash{\begin{tabular}[t]{l}\scriptsize{$\ket{r}$}\end{tabular}}}}}%
    \put(0.97737256,0.45411572){\color[rgb]{0,0,0}\rotatebox{-0.36067414}{\makebox(0,0)[lt]{\lineheight{1.25}\smash{\begin{tabular}[t]{l}\scriptsize{$\ket{e_2}$}\end{tabular}}}}}%
    \put(0.97737243,0.38401495){\color[rgb]{0,0,0}\rotatebox{-0.36067414}{\makebox(0,0)[lt]{\lineheight{1.25}\smash{\begin{tabular}[t]{l}\scriptsize{$\ket{g_2}$}\end{tabular}}}}}%
    \put(0.16608389,0.40613738){\color[rgb]{0,0,0}\rotatebox{-0.36067414}{\makebox(0,0)[lt]{\lineheight{1.25}\smash{\begin{tabular}[t]{l}\scriptsize{$\ket{s}$}\end{tabular}}}}}%
    \put(0.16608398,0.45287117){\color[rgb]{0,0,0}\rotatebox{-0.36067414}{\makebox(0,0)[lt]{\lineheight{1.25}\smash{\begin{tabular}[t]{l}\scriptsize{$\ket{g_{1}}$}\end{tabular}}}}}%
    \put(0.16608429,0.57611857){\color[rgb]{0,0,0}\rotatebox{-0.36067414}{\makebox(0,0)[lt]{\lineheight{1.25}\smash{\begin{tabular}[t]{l}\scriptsize{$\ket{e_1}$}\end{tabular}}}}}%
    \put(0,0){\includegraphics[width=\unitlength,page=2]{Setup_HOM_v4.pdf}}%
    \put(0.05564575,0.04835441){\color[rgb]{0,0,0}\makebox(0,0)[lt]{\lineheight{1.25}\smash{\begin{tabular}[t]{l}Quantum Node 1\end{tabular}}}}%
    \put(0.0551224,0.0251438){\color[rgb]{0,0,0}\makebox(0,0)[lt]{\lineheight{1.25}\smash{\begin{tabular}[t]{l}Based on DLCZ protocol\end{tabular}}}}%
    \put(0.9514698,0.04664105){\color[rgb]{0,0,0}\makebox(0,0)[rt]{\lineheight{1.25}\smash{\begin{tabular}[t]{r}Quantum Node 2\end{tabular}}}}%
    \put(0.95168363,0.02280734){\color[rgb]{0,0,0}\makebox(0,0)[rt]{\lineheight{1.25}\smash{\begin{tabular}[t]{r}Based on Rydberg Blockade\end{tabular}}}}%
    \put(0.41267925,0.04663628){\color[rgb]{0,0,0}\makebox(0,0)[lt]{\lineheight{1.25}\smash{\begin{tabular}[t]{l}Quantum interference\end{tabular}}}}%
    \put(0.34023996,0.3993631){\color[rgb]{0,0,0}\makebox(0,0)[lt]{\lineheight{1.25}\smash{\begin{tabular}[t]{l}\scriptsize{Ensemble 1}\end{tabular}}}}%
    \put(0.1461406,0.26877423){\color[rgb]{0.97647059,0.51764706,0}\makebox(0,0)[lt]{\lineheight{1.25}\smash{\begin{tabular}[t]{l}\footnotesize{photon}\end{tabular}}}}%
    \put(0.14683144,0.28654753){\color[rgb]{0.97647059,0.51764706,0}\makebox(0,0)[lt]{\lineheight{1.25}\smash{\begin{tabular}[t]{l}\footnotesize{Read}\end{tabular}}}}%
    \put(0.34396001,0.30967886){\color[rgb]{0.97647059,0.51764706,0}\makebox(0,0)[lt]{\lineheight{1.25}\smash{\begin{tabular}[t]{l}\scriptsize{pulses}\end{tabular}}}}%
    \put(0.343586,0.32443887){\color[rgb]{0.97647059,0.51764706,0}\makebox(0,0)[lt]{\lineheight{1.25}\smash{\begin{tabular}[t]{l}\scriptsize{Write}\end{tabular}}}}%
    \put(0.37199195,0.43202312){\color[rgb]{0.98823529,0.21568627,0.98823529}\makebox(0,0)[lt]{\lineheight{1.25}\smash{\begin{tabular}[t]{l}\scriptsize{photon}\end{tabular}}}}%
    \put(0.371131,0.44979642){\color[rgb]{0.98823529,0.21568627,0.98823529}\makebox(0,0)[lt]{\lineheight{1.25}\smash{\begin{tabular}[t]{l}\scriptsize{Write}\end{tabular}}}}%
    \put(0.2232078,0.4525848){\color[rgb]{0.98823529,0.21568627,0.98823529}\makebox(0,0)[lt]{\lineheight{1.25}\smash{\begin{tabular}[t]{l}\scriptsize{pulse}\end{tabular}}}}%
    \put(0.22356963,0.4703581){\color[rgb]{0.98823529,0.21568627,0.98823529}\makebox(0,0)[lt]{\lineheight{1.25}\smash{\begin{tabular}[t]{l}\scriptsize{Read}\end{tabular}}}}%
    \put(0.67154279,0.39900282){\color[rgb]{0,0,0}\makebox(0,0)[rt]{\lineheight{1.25}\smash{\begin{tabular}[t]{r}\scriptsize{Ensemble 2}\end{tabular}}}}%
    \put(0.7229485,0.20911378){\color[rgb]{0,0,0}\makebox(0,0)[rt]{\lineheight{1.25}\smash{\begin{tabular}[t]{r}\scriptsize{QWP}\end{tabular}}}}%
    \put(0.7229485,0.22718364){\color[rgb]{0,0,0}\makebox(0,0)[rt]{\lineheight{1.25}\smash{\begin{tabular}[t]{r}\scriptsize{HWP}\end{tabular}}}}%
    \put(0.72207211,0.49013113){\color[rgb]{0,0,0}\makebox(0,0)[lt]{\lineheight{1.25}\smash{\begin{tabular}[t]{l}\scriptsize{DM}\end{tabular}}}}%
    \put(0.52046575,0.22198505){\color[rgb]{0,0,0}\makebox(0,0)[lt]{\lineheight{1.25}\smash{\begin{tabular}[t]{l}\scriptsize{BS}\end{tabular}}}}%
    \put(0.43444771,0.33799362){\color[rgb]{0,0,0}\makebox(0,0)[lt]{\lineheight{1.25}\smash{\begin{tabular}[t]{l}\scriptsize{SPAD 2a}\end{tabular}}}}%
    \put(0.5183909,0.33814006){\color[rgb]{0,0,0}\makebox(0,0)[lt]{\lineheight{1.25}\smash{\begin{tabular}[t]{l}\scriptsize{SPAD 2b}\end{tabular}}}}%
    \put(0.81517252,0.26680277){\color[rgb]{0.98039216,0.17647059,0.18823529}\makebox(0,0)[lt]{\lineheight{1.25}\smash{\begin{tabular}[t]{l}\footnotesize{photon}\end{tabular}}}}%
    \put(0.81506331,0.28410404){\color[rgb]{0.98039216,0.17647059,0.18823529}\makebox(0,0)[lt]{\lineheight{1.25}\smash{\begin{tabular}[t]{l}\footnotesize{Rydberg}\end{tabular}}}}%
    \put(0,0){\includegraphics[width=\unitlength,page=3]{Setup_HOM_v4.pdf}}%
    \put(0.96587966,0.40959446){\color[rgb]{0.98039216,0.17647059,0.18823529}\makebox(0,0)[lt]{\lineheight{1.25}\smash{\begin{tabular}[t]{l}\scriptsize{photon}\end{tabular}}}}%
    \put(0.96470828,0.42601453){\color[rgb]{0.98039216,0.17647059,0.18823529}\makebox(0,0)[lt]{\lineheight{1.25}\smash{\begin{tabular}[t]{l}\scriptsize{Rydberg}\end{tabular}}}}%
    \put(0.61364146,0.25360606){\color[rgb]{0,0.6627451,0.97647059}\makebox(0,0)[lt]{\lineheight{1.25}\smash{\begin{tabular}[t]{l}\scriptsize{beam}\end{tabular}}}}%
    \put(0.61434558,0.27076084){\color[rgb]{0,0.6627451,0.97647059}\makebox(0,0)[lt]{\lineheight{1.25}\smash{\begin{tabular}[t]{l}\scriptsize{Coupling}\end{tabular}}}}%
    \put(0.96343893,0.50181301){\color[rgb]{0,0.6627451,0.97647059}\makebox(0,0)[lt]{\lineheight{1.25}\smash{\begin{tabular}[t]{l}\scriptsize{beam}\end{tabular}}}}%
    \put(0.96291292,0.51771688){\color[rgb]{0,0.6627451,0.97647059}\makebox(0,0)[lt]{\lineheight{1.25}\smash{\begin{tabular}[t]{l}\scriptsize{Coupling}\end{tabular}}}}%
    \put(0.85180613,0.50072273){\color[rgb]{0,0.6627451,0.97647059}\makebox(0,0)[rt]{\lineheight{1.25}\smash{\begin{tabular}[t]{r}\scriptsize{beam}\end{tabular}}}}%
    \put(0.85291476,0.51849589){\color[rgb]{0,0.6627451,0.97647059}\makebox(0,0)[rt]{\lineheight{1.25}\smash{\begin{tabular}[t]{r}\scriptsize{Coupling}\end{tabular}}}}%
    \put(0.77199262,0.57689665){\color[rgb]{0.97254902,0.47058824,0.9372549}\makebox(0,0)[lt]{\lineheight{1.25}\smash{\begin{tabular}[t]{l}\scriptsize{blockade}\end{tabular}}}}%
    \put(0.77112174,0.59280052){\color[rgb]{0.97254902,0.47058824,0.9372549}\makebox(0,0)[lt]{\lineheight{1.25}\smash{\begin{tabular}[t]{l}\scriptsize{Rydberg}\end{tabular}}}}%
    \put(0.74746337,0.4158472){\color[rgb]{0.97254902,0.47058824,0.9372549}\makebox(0,0)[lt]{\lineheight{1.25}\smash{\begin{tabular}[t]{l}\scriptsize{blockade}\end{tabular}}}}%
    \put(0.74659248,0.43175107){\color[rgb]{0.97254902,0.47058824,0.9372549}\makebox(0,0)[lt]{\lineheight{1.25}\smash{\begin{tabular}[t]{l}\scriptsize{Rydberg}\end{tabular}}}}%
    \put(0.62442956,0.5150531){\color[rgb]{0.98039216,0.17647059,0.18823529}\makebox(0,0)[lt]{\lineheight{1.25}\smash{\begin{tabular}[t]{l}\scriptsize{probe}\end{tabular}}}}%
    \put(0.62386525,0.52960061){\color[rgb]{0.98039216,0.17647059,0.18823529}\makebox(0,0)[lt]{\lineheight{1.25}\smash{\begin{tabular}[t]{l}\scriptsize{Input}\end{tabular}}}}%
    \put(0,0){\includegraphics[width=\unitlength,page=4]{Setup_HOM_v4.pdf}}%
    \put(0.87241153,0.40959446){\color[rgb]{0.98039216,0.17647059,0.18823529}\makebox(0,0)[lt]{\lineheight{1.25}\smash{\begin{tabular}[t]{l}\scriptsize{probe}\end{tabular}}}}%
    \put(0.87241153,0.42517248){\color[rgb]{0.98039216,0.17647059,0.18823529}\makebox(0,0)[lt]{\lineheight{1.25}\smash{\begin{tabular}[t]{l}\scriptsize{Input}\end{tabular}}}}%
    \put(0.76436946,0.30640217){\color[rgb]{0,0,0}\makebox(0,0)[lt]{\lineheight{1.25}\smash{\begin{tabular}[t]{l}\scriptsize{DM}\end{tabular}}}}%
    \put(0.28852469,0.27467919){\color[rgb]{0,0,0}\makebox(0,0)[lt]{\lineheight{1.25}\smash{\begin{tabular}[t]{l}\scriptsize{FPC 2}\end{tabular}}}}%
    \put(0.33838303,0.56297138){\color[rgb]{0,0,0}\makebox(0,0)[lt]{\lineheight{1.25}\smash{\begin{tabular}[t]{l}\scriptsize{FPC 1}\end{tabular}}}}%
    \put(0.38090834,0.50731442){\color[rgb]{0,0,0}\makebox(0,0)[lt]{\lineheight{1.25}\smash{\begin{tabular}[t]{l}\scriptsize{SPAD 1}\end{tabular}}}}%
    \put(0.14929604,0.17246063){\color[rgb]{0,0,0}\makebox(0,0)[lt]{\lineheight{1.25}\smash{\begin{tabular}[t]{l}\scriptsize{PBS}\end{tabular}}}}%
    \put(0.24446496,0.09755911){\color[rgb]{0,0,0}\makebox(0,0)[lt]{\lineheight{1.25}\smash{\begin{tabular}[t]{l}\scriptsize{AOM}\end{tabular}}}}%
    \put(0.16152144,0.51155856){\color[rgb]{0.97647059,0.51764706,0}\makebox(0,0)[lt]{\lineheight{1.25}\smash{\begin{tabular}[t]{l}\scriptsize{photon}\end{tabular}}}}%
    \put(0.16221228,0.52933186){\color[rgb]{0.97647059,0.51764706,0}\makebox(0,0)[lt]{\lineheight{1.25}\smash{\begin{tabular}[t]{l}\scriptsize{Read}\end{tabular}}}}%
    \put(0.10936738,0.59136358){\color[rgb]{0.98823529,0.21568627,0.98823529}\makebox(0,0)[lt]{\lineheight{1.25}\smash{\begin{tabular}[t]{l}\scriptsize{pulse}\end{tabular}}}}%
    \put(0.10972921,0.60913685){\color[rgb]{0.98823529,0.21568627,0.98823529}\makebox(0,0)[lt]{\lineheight{1.25}\smash{\begin{tabular}[t]{l}\scriptsize{Read}\end{tabular}}}}%
    \put(0.01292076,0.42155971){\color[rgb]{0.98823529,0.21568627,0.98823529}\makebox(0,0)[lt]{\lineheight{1.25}\smash{\begin{tabular}[t]{l}\scriptsize{photon}\end{tabular}}}}%
    \put(0.01205981,0.43933298){\color[rgb]{0.98823529,0.21568627,0.98823529}\makebox(0,0)[lt]{\lineheight{1.25}\smash{\begin{tabular}[t]{l}\scriptsize{Write}\end{tabular}}}}%
    \put(0.00220451,0.52074021){\color[rgb]{0.97647059,0.51764706,0}\makebox(0,0)[lt]{\lineheight{1.25}\smash{\begin{tabular}[t]{l}\scriptsize{pulses}\end{tabular}}}}%
    \put(0.00183049,0.53550022){\color[rgb]{0.97647059,0.51764706,0}\makebox(0,0)[lt]{\lineheight{1.25}\smash{\begin{tabular}[t]{l}\scriptsize{Write}\end{tabular}}}}%
    \put(0.24433941,0.07826322){\color[rgb]{0,0,0}\makebox(0,0)[lt]{\lineheight{1.25}\smash{\begin{tabular}[t]{l}\scriptsize{$\left( + \omega_{aom} \right)$}\end{tabular}}}}%
    \put(0,0){\includegraphics[width=\unitlength,page=5]{Setup_HOM_v4.pdf}}%
  \end{picture}%
\endgroup%

%% file: Sequence_pdflatex2.pdf_tex
\begingroup%
  \makeatletter%
  \providecommand\color[2][]{%
    \errmessage{(Inkscape) Color is used for the text in Inkscape, but the package 'color.sty' is not loaded}%
    \renewcommand\color[2][]{}%
  }%
  \providecommand\transparent[1]{%
    \errmessage{(Inkscape) Transparency is used (non-zero) for the text in Inkscape, but the package 'transparent.sty' is not loaded}%
    \renewcommand\transparent[1]{}%
  }%
  \providecommand\rotatebox[2]{#2}%
  \newcommand*\fsize{\dimexpr\f@size pt\relax}%
  \newcommand*\lineheight[1]{\fontsize{\fsize}{#1\fsize}\selectfont}%
  \ifx\svgwidth\undefined%
    \setlength{\unitlength}{245.2971547bp}%
    \ifx\svgscale\undefined%
      \relax%
    \else%
      \setlength{\unitlength}{\unitlength * \real{\svgscale}}%
    \fi%
  \else%
    \setlength{\unitlength}{\svgwidth}%
  \fi%
  \global\let\svgwidth\undefined%
  \global\let\svgscale\undefined%
  \makeatother%
  \begin{picture}(1,0.53138155)%
    \lineheight{1}%
    \setlength\tabcolsep{0pt}%
    \put(0,0){\includegraphics[width=\unitlength,page=1]{Sequence_pdflatex2.pdf}}%
    \put(0.45658766,0.41936604){\color[rgb]{0,0,0}\makebox(0,0)[rt]{\lineheight{1.25}\smash{\begin{tabular}[t]{r}\small{SPAD1}\end{tabular}}}}%
    \put(0,0){\includegraphics[width=\unitlength,page=2]{Sequence_pdflatex2.pdf}}%
    \put(0.05259967,0.20890031){\color[rgb]{0,0,0}\makebox(0,0)[rt]{\lineheight{1.25}\smash{\begin{tabular}[t]{r}Node 2\end{tabular}}}}%
    \put(0.05259967,0.45415374){\color[rgb]{0,0,0}\makebox(0,0)[rt]{\lineheight{1.25}\smash{\begin{tabular}[t]{r}Node 1\end{tabular}}}}%
    \put(0,0){\includegraphics[width=\unitlength,page=3]{Sequence_pdflatex2.pdf}}%
    \put(0.69724392,0.47026053){\color[rgb]{0,0,0}\makebox(0,0)[rt]{\lineheight{1.25}\smash{\begin{tabular}[t]{r}\scriptsize{Storage time}\end{tabular}}}}%
    \put(0.52867346,0.19361875){\color[rgb]{0,0,0}\makebox(0,0)[lt]{\lineheight{1.25}\smash{\begin{tabular}[t]{l}\textbf{\scriptsize{Distinguishable case}}\end{tabular}}}}%
    \put(0.21530354,0.49940958){\color[rgb]{0.94117647,0.30588235,0.30588235}\makebox(0,0)[rt]{\lineheight{1.25}\smash{\begin{tabular}[t]{r}\scriptsize{Writing pulses}\end{tabular}}}}%
    \put(0.30007689,0.49940958){\color[rgb]{0.90980392,0.57254902,0.1254902}\makebox(0,0)[lt]{\lineheight{1.25}\smash{\begin{tabular}[t]{l}\scriptsize{Write photon detection}\end{tabular}}}}%
    \put(0.95106077,0.49940958){\color[rgb]{0.94117647,0.30588235,0.30588235}\makebox(0,0)[rt]{\lineheight{1.25}\smash{\begin{tabular}[t]{r}\scriptsize{Photon}\end{tabular}}}}%
    \put(0,0){\includegraphics[width=\unitlength,page=4]{Sequence_pdflatex2.pdf}}%
    \put(0.70775464,0.27756129){\color[rgb]{0,0,0}\makebox(0,0)[rt]{\lineheight{1.25}\smash{\begin{tabular}[t]{r}\scriptsize{Synchronization}\end{tabular}}}}%
    \put(0,0){\includegraphics[width=\unitlength,page=5]{Sequence_pdflatex2.pdf}}%
    \put(0.53260706,0.00954514){\color[rgb]{0,0,0}\makebox(0,0)[rt]{\lineheight{1.25}\smash{\begin{tabular}[t]{r}\scriptsize{Time}\end{tabular}}}}%
    \put(0.52867346,0.34777814){\color[rgb]{0,0,0}\makebox(0,0)[lt]{\lineheight{1.25}\smash{\begin{tabular}[t]{l}\textbf{\scriptsize{Indistinguishable case}}\end{tabular}}}}%
    \put(0.95106078,0.48539509){\color[rgb]{0.94117647,0.30588235,0.30588235}\makebox(0,0)[rt]{\lineheight{1.25}\smash{\begin{tabular}[t]{r}\scriptsize{generation}\end{tabular}}}}%
    \put(0,0){\includegraphics[width=\unitlength,page=6]{Sequence_pdflatex2.pdf}}%
    \put(0.31830563,0.25795199){\color[rgb]{0.58823529,0,0.61960784}\makebox(0,0)[rt]{\lineheight{1.25}\smash{\begin{tabular}[t]{r}\scriptsize{Dipole trapping}\end{tabular}}}}%
    \put(0.26673421,0.17947373){\color[rgb]{0.58823529,0,0.61960784}\makebox(0,0)[rt]{\lineheight{1.25}\smash{\begin{tabular}[t]{r}\scriptsize{ON}\end{tabular}}}}%
    \put(0.46881595,0.17835262){\color[rgb]{0.58823529,0,0.61960784}\makebox(0,0)[rt]{\lineheight{1.25}\smash{\begin{tabular}[t]{r}\scriptsize{OFF}\end{tabular}}}}%
    \put(0,0){\includegraphics[width=\unitlength,page=7]{Sequence_pdflatex2.pdf}}%
    \put(0.41401738,0.34539667){\color[rgb]{0,0,0}\makebox(0,0)[rt]{\lineheight{1.25}\smash{\begin{tabular}[t]{r}\scriptsize{Classical}\\\scriptsize{signal}\end{tabular}}}}%
    \put(0.70775464,0.25310118){\color[rgb]{0,0,0}\makebox(0,0)[rt]{\lineheight{1.25}\smash{\begin{tabular}[t]{r}\scriptsize{time}\end{tabular}}}}%
  \end{picture}%
\endgroup%